# Motion Equation of Vorticity for Newton Fluid

*Xiao Jianhua*

*Natural Science Foundation Research Group, Shanghai Jiaotong University*

**Abstract:** The vorticity plays an important role in aerodynamics and rotational flow. Usually, they are studied with modified Navier-Stokes equation. This research will deduce the motion equation of vorticity from Navier-Stokes equation. To this propose, the velocity gradient field is decomposed as the stack of non-rotation field and pure-rotation field. By introducing the Chen's S+R decomposition, the rotational flow is redefined. For elastic fluid, the research shows that for Newton fluid, the local average rotation always produces an additional pressure on the rotation plane. This item is deterministic rather than stochastic (as Reynolds stress) or adjustable. For non-elastic fluid, such as air, the research shows that the rotation will produce an additional stress along the rotation axis direction, that is on the normal direction of rotation plane. This result can be used to explain the lift force connected with vortex. The main purpose of this research is to supply a solvable mathematical model for the calculation of vorticity and pressure when suitable boundary condition is adapted. Based on this understanding, some way to control the movement of vortices may be produced.

## 1. Introduction

The Navier-Stokes equations are widely used in fluid mechanics as basic equations. Its effectiveness for simple fluid motion is evident. In fact, Navier-Stokes equation has been unqualifiedly used as the unique foundation of whole fluid dynamics. But, recently, enquiry about its reasonability and solvability has been proposed by many researchers [1-2]. Some even goes to claim that the turbulence cannot be solved by the Navier-Stokes equation [3]. Then, as a natural consequence, we want to answer the enquiry about its solvability.

Firstly, the introduction of motion equation is purely based the analogy of solid continuum. As this equation is "artificially" enforced on the motion of fluid, the pressure in Navier-Stokes equation has lost its intrinsic meaning as "static pressure" and become an adjustable "parameter" in the calculation of fluid mechanics. These phenomena are common practice in the fluid field calculation [4-6].

Secondly, the definition of vortex, based on the Stokes's S+R decomposition [7], cannot well describe the local rotation and global rotation of fluid, when the average local rotation is the main role. However, for small average local rotation, it is a good approximation [7]. Hence, one can only say that the vortex is ill-defined. The ill-definition of vortex is the main cause which misleads many researchers on rotational flows.

As the consequence of above mentioned problems, the Navier-Stokes equation is ill-used in fluid mechanics. Based on this research, the Navier-Stokes equation is correct in form. The research shows that, by discarding the misunderstanding and extending the Stokes's S+R decomposition to Chen's S+R decomposition [7], the Navier-Stokes equation can be used for complicated flow and for turbulence. The results may help to rebuild our confidence about the Navier-Stokes equation for complicated flow and turbulence in fluid mechanics

This paper, firstly, reformulates the Navier-Stokes equation by Chen's S+R decomposition and, hence, reinterprets its intrinsic meaning. Then, the paper goes to define the "vortex", based on the Chen's S+R decomposition. By this way, the Navier-Stokes equation is decomposed into two kinds of motion equations. One is for non-rotation flow which is the same as the traditional form used in practice, another is for rotational flow which gives the motion equation for the "vortex" defined in the paper. For comparing, the intrinsic meaning of incompressible fluid is discussed, which is only meaningful for elastic flow.

## 2. Decomposition of Navier-Stokes Equation

In standard rectangular coordinator system, the Navier-Stokes equation widely used in fluid mechanics as basic equations is:

$$\frac{\partial}{\partial t}(\rho u^i) = -\rho u^j \frac{\partial u^i}{\partial x^j} - \frac{\partial p_0}{\partial x^j}\delta^i_j + \lambda \frac{\partial}{\partial x^j}(\frac{\partial u^l}{\partial x^l})\delta^i_j + 2\mu \frac{\partial}{\partial x^j}(\frac{\partial u^i}{\partial x^j}) \tag{1}$$

For small rotational fluid motion, the Stokes's S+R decomposition for velocity gradient is:

$$\frac{\partial u^i}{\partial x^j} = \frac{1}{2}(\frac{\partial u^i}{\partial x^j} + \frac{\partial u^j}{\partial x^i}) + \frac{1}{2}(\frac{\partial u^i}{\partial x^j} - \frac{\partial u^j}{\partial x^i}) = s^i_j + w^i_j \tag{2}$$

is widely used as a standard decomposition method. Here, the upper index and lower index is used to show the non-commutability. Its strict mathematics theory will not be discussed in this paper. Hence, it can be taken as a form for convenience. Put this equation into equation (1), one will get:

$$\frac{\partial}{\partial t}(\rho u^i) = -\rho u^l (s^i_j + w^i_j) - \frac{\partial p_0}{\partial x^i} + \lambda \frac{\partial}{\partial x^j}(s^l_l + w^l_l)\delta^i_j + 2\mu \frac{\partial}{\partial x^j}(s^i_j + w^i_j) \tag{3}$$

where, the $w^i_j$ is related with infinitesimal local rotation. For larger rotation, it is well known that the $w^i_j$ cannot represent the correct local rotation.

For much larger rotation, when the condition:

$$-w^i_j w^j_i < 2 \tag{4}$$

is met, it can be decomposed as Chen's S+R form [7] as:

$$\frac{\partial u^i}{\partial x^j} = S^i_j + R^i_j - \delta^i_j \tag{5}$$

where:

$$S^i_j = \frac{1}{2}(\frac{\partial u^i}{\partial x^j} + \frac{\partial u^j}{\partial x^i}) - (1 - \cos \Theta) L^i_k L^k_j \tag{6}$$

$$R^i_j = \delta^i_j + \sin \Theta \cdot L^i_j + (1 - \cos \Theta) L^i_k L^k_j \tag{7}$$

$$L^i_j = \frac{1}{2 \sin \Theta}(\frac{\partial u^i}{\partial x^j} - \frac{\partial u^j}{\partial x^i}) = \frac{1}{\sin \Theta} w^i_j \tag{8}$$

$$\sin \Theta = (-\frac{1}{2} w^i_j w^j_i)^{\frac{1}{2}} \tag{9}$$

The parameter $\Theta$ represents local average rotation and its value range is $[0, \frac{\pi}{2})$, $L^i_j$ represents the local average rotation direction tensor, $R^i_j$ is an unit-orthogonal rotation tensor.

Hence, when the condition (4) is met, the Navier-Stokes equation (1) can be decomposed as:

$$\frac{\partial}{\partial t}(\rho u^i) = -\rho u^l (S^i_j + R^i_j - \delta^i_j) - \frac{\partial p_0}{\partial x^i} + \lambda \frac{\partial}{\partial x^j}(S^l_l + R^l_l)\delta^i_j + 2\mu \frac{\partial}{\partial x^j}(S^i_j + R^i_j) \tag{10}$$

When the condition (4) is not met, based on Chen's S+R decomposition, the velocity gradient still can be decomposed as, in form [8]:

$$\frac{\partial u^i}{\partial x^j} = \widetilde{S}^i_j + \frac{1}{\cos \theta} \widetilde{R}^i_j - \delta^i_j \tag{11}$$

Where:

$$\widetilde{S}^i_j = \frac{1}{2}(u^i|_j + u^i|_j{}^T) - (\frac{1}{\cos \theta} - 1)(\widetilde{L}^i_k \widetilde{L}^k_j + \delta^i_j) \tag{12}$$

$$(\cos\theta)^{-1}\tilde{R}^i_j = \delta^i_j + \frac{\sin\theta}{\cos\theta}\tilde{L}^i_j + (\frac{1}{\cos\theta}-1)(\tilde{L}^i_k\tilde{L}^k_j + \delta^i_j) \tag{13}$$

$$\tilde{R}^i_j = \delta^i_j + \sin\theta \cdot \tilde{L}^i_j + (1-\cos\theta)\tilde{L}^i_k\tilde{L}^k_j \tag{14}$$

$$\tilde{L}^i_j = \frac{\cos\theta}{2\sin\theta}(u^i|_j - u^i|_j^T) \tag{15}$$

$$(\cos\theta)^{-2} = 1 + \frac{1}{4}[(u^1|_2 - u^2|_1)^2 + (u^2|_3 - u^3|_2)^2 + (u^3|_1 - u^1|_3)^2] \tag{16}$$

The parameter $\theta$ represents local average rotation and its value range is $[0,\frac{\pi}{2})$, $\tilde{L}^i_j$ represents the local average rotation direction tensor, $\tilde{R}^i_j$ is an unit-orthogonal rotation tensor.

So, from equations (5) and (11), there are two kinds of fluid rotation motion. For the first form, i.e. equation (5), when $S^i_j = 0$ the velocity of fluid has no amplitude variation but has direction variation, which is expresses by local rotation. For the second form, i.e. equation (11), when $\tilde{S}^i_j = 0$ the velocity of fluid not only has a pure local rotation but also has an amplitude incremental $\frac{1}{\cos\theta}-1$.

As the Navier-Stokes equation is linear in form, any fluid motion can be decomposed as the stack of a non-rotational flow and a pure rotational flow. For the non-rotational has been well studied, this paper will put focus on the pure rotational flow.

## 3. Rotational Flow with Chen's Decomposition Form-One

For the fluid motion, expressed by equation (5), there are two typical cases.

(1). Non-rotation flow, defined by $R^i_j = \delta^i_j$. For non-rotational flow, the intrinsic strain rate is identical with the traditional form:

$$S^i_j = \frac{1}{2}(\frac{\partial u^i}{\partial x^j} + \frac{\partial u^j}{\partial x^i}) \equiv s^i_j \tag{17}$$

The Navier-Stokes equation is in traditional form for Newton fluid:

$$\frac{\partial}{\partial t}(\rho u^i) = -\rho u^j s^i_j - \frac{\partial p_0}{\partial x^j}\delta^i_j + \lambda\frac{\partial}{\partial x^j}(s^l_l)\delta^i_j + 2\mu\frac{\partial}{\partial x^j}(s^i_j) \tag{18}$$

(2). Pure rotation flow, defined by $S^i_j = 0$. For such a kind of flow, the Navier-Stokes equation (1) gives out:

$$\frac{\partial}{\partial t}(\rho u^i) = -\rho u^j(R^i_j - \delta^i_j) - \frac{\partial p_0}{\partial x^j}\delta^i_j + \lambda\frac{\partial}{\partial x^j}(R^l_l - 3)\delta^i_j + 2\mu\frac{\partial}{\partial x^j}(R^i_j) \tag{19}$$

Only when $\Theta$ is infinitesimal, that is when one has approximation $(1-\cos\Theta) \approx 0$, on can get:

$$\frac{\partial}{\partial t}(\rho u^i) \approx -\rho u^j w^i_j - \frac{\partial p_0}{\partial x^i} + \mu\frac{\partial}{\partial x^j}w^i_j \tag{20}$$

This shows that for pure infinitesimal local rotation the conventional Navier-Stokes equation still is a good approximation, although its physical foundation is misunderstood.

Based on Chen's S+R decomposition (6), when $S^i_j = 0$, one will have:

$$s^i_j = \frac{1}{2}(\frac{\partial u^i}{\partial x^j} + \frac{\partial u^j}{\partial x^i}) = (1-\cos\Theta)L^i_k L^k_j \tag{21}$$

It shows that for a pure local rotation flow, the traditional strain rate $s^i_j$ is determined by the local rotation axe direction and local average rotation angle $\Theta$. It is completely determined by three parameters. From this point to see, a pure local rotation flow should not be described by the vortex defined by the curl of velocity field. The more suitable definition of pure rotation should be equation (21) or $S^i_j = 0$.

Notes that for plane rotation, if the rotation axe is along the $x^3$ direction, the equation (21) becomes:

$$s^i_j = (1-\cos\Theta)L^i_k L^k_j = (1-\cos\Theta)\begin{vmatrix} 1 & 0 & 0 \\ 0 & 1 & 0 \\ 0 & 0 & 0 \end{vmatrix} \quad (22)$$

That is to say that a pure plane local rotation will cause rotation plane symmetric velocity gradient. In this case, if one uses traditional strain rate definition to calculate stress, he will report that there is an additional plane pressure. As we know that such an additional plane pressure does not exist for pure local rotation flow, one may find that symmetric strain rate should be defined by $S^i_j$ rather than $s^i_j$.

To see the $S^i_j = 0$ indeed defines the pure rotation motion, let write down the velocity field variation in full form:

$$u^i = (\frac{\partial u^i}{\partial x^j} + \delta^i_j)U^j = R^i_j U^j \quad (23)$$

where, $U^j$ represents the velocity of a neighboring reference space point. Hence:

$$u^i u^i = (R^i_k U^k)(R^i_l U^l) = R^i_k R^i_l U^k U^l = U^k U^k \quad (24)$$

It shows that the velocity field is a pure rotation field. In this sense, it can be called as elastic rotation.

So, the pure rotation of fluid should be defined by $R^i_j$ for its rotation axe direction instead of the definition of vortex and by $\Theta$ for its magnitude instead of vortices. For the pure rotation flow, the $L^i_j$ has only two independent parameters. So, if the mean flow velocity $U^i$ is given, equations (19) and (23) form a closed motion equation, where the pure static pressure $p_0$ is viewed as the intrinsic parameter of fluid. That is:

$$\frac{\partial}{\partial t}(\rho R^i_j U^j) = -\rho R^j_l U^l(R^i_j - \delta^i_j) - \frac{\partial p_0}{\partial x^j}\delta^i_j + \lambda\frac{\partial}{\partial x^j}(R^l_l - 3)\delta^i_j + 2\mu\frac{\partial}{\partial x^j}(R^i_j) \quad (25)$$

As the $R^i_j$ is an unit orthogonal tensor, the equation can be simplified as:

$$\frac{\partial}{\partial t}(\rho R^i_j U^j) = \rho(R^i_l U^l - U^i) - \frac{\partial p_0}{\partial x^j}\delta^i_j - 2\lambda\frac{\partial \cos\Theta}{\partial x^j}\delta^i_j + 2\mu\frac{\partial}{\partial x^j}(R^i_j) \quad (26)$$

It is the tensor form of angular momentum conservation equation, corresponds to the usual Euler's momentum equation. In conventional fluid mechanics, it corresponds to the vorticity transport equation. In form, it is a highly non-linear equation, some unstable solution may exist. So, this equation can be used to predict turbulent flow. To make this problem clear, the incompressible flow is considered below.

## 4. Incompressible Flow in Chen's Decomposition Form-One

Traditionally, the incompressible flow is defined by:

$$\frac{\partial u^i}{\partial x^i} = 0 \tag{27}$$

Appling this equation to Chen's S+R decomposition form-one, one will get:

$$S_i^i + R_i^i - 3 = 0 \tag{28}$$

By equation (7), one has:

$$R_i^i - 3 = 2(1 - \cos\Theta) \tag{29}$$

Hence, the incompressible flow is defined by:

$$S_i^i = -2(1 - \cos\Theta) \tag{30}$$

It shows that not only the $S_j^i$ can be completely determined by parameters of orthogonal rotation tensor.

Based on this research, the incompressible flow can be divided into the addition of a pure rotation and a volume contraction caused by local average rotation. Hence, the traditional definition of impressible flow is not correct when the local average rotation angle is bigger enough.

For incompressible Newton fluid, the reformulated Navier-Stokes equation is:

$$\frac{\partial}{\partial t}(\rho u^i) = -\rho u^l (S_j^i + R_j^i - \delta_j^i) - \frac{\partial p_0}{\partial x^i} + 2\mu \frac{\partial}{\partial x^j}(S_j^i + R_j^i) \tag{31}$$

Comparing with equation (19), if the equation (19) is satisfied, to satisfy equation (31), one gets an additional equation for $S_j^i$:

$$-\rho u^j (S_j^i) - \lambda \frac{\partial}{\partial x^j}(R_l^l - 3)\delta_j^i + 2\mu \frac{\partial}{\partial x^j}(S_j^i) = 0 \tag{32}$$

That is:

$$-\rho u^j (S_j^i) + 2\lambda \frac{\partial \cos\Theta}{\partial x^j}\delta_j^i + 2\mu \frac{\partial}{\partial x^j}(S_j^i) = 0 \tag{33}$$

Hence, for traditional incompressible flow, the equations (26) and (33) form closed-form equations. Therefore, traditional incompressible flow is solvable mathematically. The striking feature of equations (26) and (33) is that it shows the parameter $\lambda$ has effects on traditional incompressible flow when the local rotation angular $\Theta$ is big enough. This conclusion has been well discovered by experiments.

## 5. Rotational Flow with Chen's Decomposition Form-Two

The pure rotation flow in Chen's decomposition form-two is defined by $\tilde{S}_j^i = 0$. For such a kind of flow, the Navier-Stokes equation (1) gives out:

$$\frac{\partial}{\partial t}(\rho u^i) = -\rho u^j (\frac{1}{\cos\theta}\tilde{R}_j^i - \delta_j^i) - \frac{\partial p_0}{\partial x^j}\delta_j^i + \lambda \frac{\partial}{\partial x^j}(\frac{1}{\cos\theta}\tilde{R}_l^l - 3)\delta_j^i + 2\mu \frac{\partial}{\partial x^j}(\frac{1}{\cos\theta}\tilde{R}_j^i) \tag{34}$$

Based on Chen's S+R decomposition (12), when $\tilde{S}_j^i = 0$, one will have:

$$s_j^i = \frac{1}{2}(\frac{\partial u^i}{\partial x^j} + \frac{\partial u^j}{\partial x^i}) = (\frac{1}{\cos\theta} - 1)(\tilde{L}_k^i \tilde{L}_j^k + \delta_j^i) \tag{35}$$

It shows that for a pure local rotation flow, the traditional strain rate $s_j^i$ is determined by the local rotation axe direction and local average rotation angle $\theta$. Notes that for plane rotation, if the rotation axe is along the $x^3$ direction, the equation (35) becomes:

$$s^i_j = (\frac{1}{\cos\theta} - 1)(\tilde{L}^i_k \tilde{L}^k_j + \delta^i_j) = (\frac{1}{\cos\theta} - 1)\begin{vmatrix} 0 & 0 & 0 \\ 0 & 0 & 0 \\ 0 & 0 & 1 \end{vmatrix} \tag{36}$$

That is to say that a pure plane local rotation will not cause rotation plane symmetric velocity gradient. In contrast, it cause the velocity increase along the rotational axis direction. For air flow, it expresses the hurricane. For water flow, it expresses the water burst along the normal of rotation plane. In fact, it is this mechanism which supplies the lift force needed for air-plane flying.

To see the velocity variation for $\tilde{S}^i_j = 0$, let write down the velocity field variation in full form:

$$u^i = (\frac{\partial u^i}{\partial x^j} + \delta^i_j)U^j = \frac{1}{\cos\theta}\tilde{R}^i_j U^j \tag{37}$$

where, $U^j$ represents the velocity of a neighboring reference space point. Hence:

$$u^i u^i = (\frac{1}{\cos\theta}R^i_k U^k)(\frac{1}{\cos\theta}R^i_l U^l) = (\frac{1}{\cos\theta})^2 R^i_k R^i_l U^k U^l = (\frac{1}{\cos\theta})^2 U^k U^k \tag{38}$$

It shows that the velocity field is increased by a factor $1/\cos\theta$. Physically, it shows that the kinetic energy increase of fluid is achieved by such a form of rotation. For air-plane flying, the high the speed of flying is the bigger the lift force is [in fact the static lift stress is $(\lambda + 2\mu)(\frac{1}{\cos\theta} - 1)$ ].

So, if the mean flow velocity $U^i$ is given, equations (34) and (37) form a closed motion equation, where the pure static pressure $p_0$ is viewed as the intrinsic parameter of fluid. That is:

$$\frac{\partial}{\partial t}(\frac{\rho}{\cos\theta}\tilde{R}^i_j U^j) = -\frac{\rho}{\cos\theta}\tilde{R}^i_l U^l(\frac{1}{\cos\theta}\tilde{R}^i_j - \delta^i_j) - \frac{\partial p_0}{\partial x^j}\delta^i_j + \lambda\frac{\partial}{\partial x^j}(\frac{1}{\cos\theta}\tilde{R}^l_l - 3)\delta^i_j + 2\mu\frac{\partial}{\partial x^j}(\frac{1}{\cos\theta}\tilde{R}^i_j) \tag{39}$$

As the $\tilde{R}^i_j$ is an unit orthogonal tensor, the equation can be simplified as:

$$\frac{\partial}{\partial t}(\frac{\rho}{\cos\theta}\tilde{R}^i_j U^j) = -\frac{\rho}{(\cos\theta)^2}U^i + \frac{\rho}{\cos\theta}\tilde{R}^i_j U^j - \frac{\partial p_0}{\partial x^j}\delta^i_j + 2\lambda\frac{\partial}{\partial x^j}(\frac{1}{\cos\theta})\delta^i_j + 2\mu\frac{\partial}{\partial x^j}(\tilde{R}^i_j) \tag{40}$$

It is the tensor form of angular momentum conservation equation, corresponds to the usual Euler's momentum equation. In conventional fluid mechanics, it corresponds to the vorticity transport equation. In form, it is a highly non-linear equation, some unstable solution may exist. So, this equation can be used to predict turbulent flow.

If one takes the $U^i$ as the air plane speed, omitting the space variation, the striking feature of equation (40) is that the equation shows the dynamic air-plane lift force can be approximated as:

$$F^i_{lift} = \frac{\rho}{\cos\theta}\tilde{R}^i_j U^j \tag{41}$$

The lift is just the Kutta-Joukouski theorem[9] in air-plane co-moving coordinator system.

Based above discussion, it is reasonable to see that the equation (40) can be used to study aerodynamics.

## 6. Further Discussion

The research presented here is focused on the rotational flow described by Navier-Stokes equation. By introducing the Chen's S+R decomposition, the rotational flow is redefined. For elastic fluid, the Chen's decomposition form-one should be used. In this case, the research shows that for Newton fluid, the local average rotation always produces an additional pressure on the rotation plane. This item is deterministic rather than stochastic (as Reynolds stress) or adjustable. For non-elastic fluid, such as air, the Chen's decomposition form-two should be used. For this case, the research shows that the rotation will produce an additional stress along the rotation axis direction, that is on the normal direction of rotation plane. This result can be used to explain the lift force connected with vortex. The research shows that the vortices of fluid have two intrinsic different forms, which corresponds to the two possible forms of Chen's decomposition of velocity gradient. This is very important to understand the many controversial results in fluid mechanics. For the transition condition from one-form to two-form or its reserve will not be discussed in this paper, as these problems should be related with the feature of fluid. The main purpose of this research is to supply a solvable mathematical model for the calculation of vorticity and pressure when suitable boundary condition is adapted.